\input harvmac
\input epsf
\def\tb{\tan\beta}
\def\ra{\rightarrow}

\def\gsim{{~\raise.15em\hbox{$>$}\kern-.85em
          \lower.35em\hbox{$\sim$}~}}
\def\lsim{{~\raise.15em\hbox{$<$}\kern-.85em
          \lower.35em\hbox{$\sim$}~}}

%%%%%%%%%%%%%%%%%%%%%%%%%%%%%%%%%%%%%%%%
\Title{hep-ph/9607397, SCIPP 96/30, WIS-96/29/Jul-PH}
{\vbox{\centerline{Variations on Minimal}
\centerline{Gauge Mediated Supersymmetry Breaking}}}
\bigskip
\centerline{Michael Dine$^a$, Yosef Nir$^b$ and Yuri Shirman$^a$}
\smallskip
\centerline{\it $^a$Santa Cruz Institute for Particle Physics,
 University of California, Santa Cruz, CA 95064}
\centerline{\it $^b$Department of Particle Physics,
 Weizmann Institute of Science, Rehovot 76100, Israel}
\bigskip
\bigskip
\baselineskip 18pt
 
\noindent
We study various modifications to the minimal models of gauge mediated
supersymmetry breaking. We argue that, under reasonable assumptions, the
structure of the messenger sector is rather restricted. We investigate
the effects of possible mixing between messenger and ordinary
squark and slepton fields and, in particular, violation of universality.
We show that acceptable values for the $\mu$ and $B$ parameters can
naturally arise from discrete, possibly horizontal, symmetries.
We claim that in models where the supersymmetry breaking parameters
$A$ and $B$
vanish at tree level, $\tb$ could be large without fine tuning.
We explain how the supersymmetric CP problem is solved in such models.
 
\Date{7/96}
%\draftmode

%%%%%%%%%%%%%%%%%%%%%
%%%%%%%%%%%%%%%%%%%%%
\newsec{Introduction}
 
Most speculations about supersymmetry phenomenology start
with the assumption that
supersymmetry is broken at an extremely large energy scale,
of order $10^{11}$ GeV, and that the breaking is fed
down to the partners of ordinary fields through
gravitational interactions.  There has been renewed
interest, recently, in the possibility that supersymmetry
might be broken at much lower energies, of order
$10$'s to $100$'s of TeV.  This interest has grown
out of an appreciation of the supersymmetric flavor problem,
as well as out of successful efforts to build models
with dynamical supersymmetry breaking at low energies
\nref\dns{M. Dine, A.E. Nelson and Y. Shirman,
Phys. Rev. {\bf D51} (1995) 1362, hep-ph/9408384.}%
\nref\dnns{M. Dine, A.E. Nelson, Y. Nir and Y. Shirman,
Phys. Rev. {\bf D53} (1996) 2658, hep-ph/9507378.}%
\refs{\dns-\dnns}. More recently, it has also been fueled by one small
piece of experimental support: a single $e^+e^-\gamma\gamma\not E_T$
event observed at CDF
\nref\cdf{S. Park, representing the CDF Collaboration, in {\it 10th
Topical Workshop on Proton-Antiproton Collider Physics}, eds.
R. Raha and J. Yoh (AIP Press, New York, 1995).}%
\nref\ddrt{S. Dimopoulos, M. Dine, S. Raby and S. Thomas,
Phys. Rev. Lett. 76 (1996) 3494, hep-ph/9601367.}%
\nref\kaneetalone{S. Ambrosanio, G.L. Kane, G.D. Kribs, S.P. Martin
and S. Mrenna, Phys. Rev. Lett. 76 (1996) 3498, hep-ph/9602239.}%
\nref\cpetal{D.R. Stump, M. Wiest and C.P. Yuan, {\it Detecting a Light
Gravitino at Linear Collider to Probe the SUSY Breaking
Scale,} hep-ph/9601362}%
\nref\dtw{S. Dimopoulos, S. Thomas and J.D. Wells, {\it Implications of
Low Energy Supersymmetry Breaking at the Tevatron},
SLAC-PUB-7148, hep-ph/9604452.}%
\nref\kaneetaltwo{S. Ambrosanio, G.L. Kane, G.D. Kribs, S.P. Martin
and S. Mrenna, {\it Search for supersymmetry with a Light Gravitino at
the Fermilab Tevatron and CERN LEP Colliders}, hep-ph/9605398.}%
\refs{\cdf-\kaneetaltwo}.
 
Existing models of low energy supersymmetry breaking assume that gauge
interactions are the messengers of supersymmetry breaking. This mechanism
is referred to as ``gauge mediated supersymmetry breaking'' (GMSB). Such
models are highly predictive. Indeed, all $106$ new parameters of the
minimal supersymmetric standard model are typically predicted in terms of
two or three new parameters.  For example, the simplest model (so-called
``Minimal Gauge Mediation," or MGM) possesses a messenger sector
consisting of a single $5+\bar 5$ of $SU(5)$, i.e. color triplets,
$q+\bar q$, and weak doublets, $\ell+\bar \ell$. These couple to a single
gauge singlet field, $S$, through a superpotential,
\eqn\simplemodel{W=\lambda_1 S q \bar q + \lambda_2 S \ell \bar \ell.}
The field $S$ has a non-zero expectation value both for its scalar and
auxiliary components, $S$ and $F_S$. Integrating out the messenger sector
gives rise to gaugino masses at one loop, and scalar masses
at two loops. For the gauginos, one has:
\eqn\gauginomasses{m_{\lambda_i}=c_i\ {\alpha_i\over4\pi}\ \Lambda\ ,}
where $\Lambda = F_S/S$, $c_1=5/3$, $c_2=c_3=1$, and $\alpha_1=
\alpha/\cos^2\theta_W$. For the scalar masses one has:
\eqn\scalarmasses{\tilde m^2 ={2 \Lambda^2}
\left[C_3\left({\alpha_3 \over 4 \pi}\right)^2
+C_2\left({\alpha_2\over 4 \pi}\right)^2
+{5 \over 3}{\left(Y\over2\right)^2}
\left({\alpha_1\over 4 \pi}\right)^2\right],}
where $C_3 = 4/3$ for color triplets and zero for singlets,
$C_2= 3/4$ for weak doublets and zero for singlets,
and $Y=2(Q-T_3)$ is the ordinary hypercharge.\foot{These formulas
predict a near degeneracy of the bino and the right handed sleptons.
Important corrections due to operator renormalization and $D$ terms have
been discussed in
\ref\bkw{K.S. Babu, C. Kolda and F. Wilczek, {\it Experimental
Consequences of a Minimal Messenger Model for Supersymmetry Breaking},
IASSNS-HEP 96/55, hep-ph/9605408.}.}
 
Because the scalar masses are functions of only gauge quantum numbers,
these models also automatically solve the supersymmetric flavor problem.
This feature is preserved in any theory in which gauge interactions are
the messengers of supersymmetry breaking. As a result, such models don't
suffer from flavor changing neutral currents, and
can naturally have small CP violation.
 
One can argue, based on these features alone, that low energy
supersymmetry breaking is in many ways more appealing than models with
intermediate scale breaking.  In fact, it is fair to say there do not
yet exist computable models of intermediate scale breaking.\foot{
Supergravity models are non-renormalizable, so without some underlying
finite theory, none of the soft breakings can be determined, even if
the supersymmetry breaking
mechanism  is specified.  In string theory, one cannot compute the
soft breaking terms without understanding the dynamics which selects a
particular point in the moduli space. While regions of the moduli space
in which string theory might yield squark degeneracy have been identified
\nref\ibanezlust{L.E. Ibanez and D. Lust,
 Nucl. Phys. {\bf B382} (1992) 305.}%
\nref\kl{V. Kaplunovsky and J. Louis, Phys. Lett. B306 (1993) 269.}%
\nref\bdmtheory{T. Banks and M. Dine, {\it Couplings and Scales in
Strongly Coupled Heterotic String Theory}, RU-96-27, hep-th/9605136..}%
\refs{\ibanezlust-\bdmtheory}, it is difficult to
understand why these regions would be preferred.}
On the other hand, while successful models of low energy
breaking have been constructed, it would be difficult
to claim that we are yet in possession of the analog
of the Weinberg-Salam model for supersymmetry, a model
compelling for its elegance and simplicity.
 
The mass formulae of eqns. \gauginomasses\ and \scalarmasses\ are
remarkably predictive. But given that we do not yet possess a compelling
model, it is natural to ask in what sense such formulas are inevitable
consequences of low energy supersymmetry breaking. In ref. \ddrt, some
plausible modifications of these formulas were mentioned. In this paper,
we will attempt a more systematic analysis of this issue. In section 2,
we will consider weakly coupled models. In such theories, some very
modest assumptions severely restrict the allowed possibilities. Gauge
mediation must play a dominant role, and the messenger sector must
consist of small numbers of vectorlike representations of $SU(5)$.
As already discussed in ref. \ddrt, the overall coefficients in eqns.
\scalarmasses\ and \gauginomasses\ may change.
 
But we also find that it is possible to obtain departures from
universality.\foot{In this paper, we will use the term ``universality''
to mean scalar masses which are functions only of gauge quantum
numbers, and $A$ terms which are small or proportional to
fermion Yukawa couplings.}
 In particular, in models such as those of ref. \dnns, it
has been assumed that the messenger sector is completely separated from
the visible sector. This can be assured by discrete symmetries. However,
one can consider relaxing this condition. Indeed, one might well want
to since otherwise the models possess stable particles which are
problematic in cosmology.\foot{In
\ref\dgp{S. Dimopoulos, G.F. Giudice and A. Pomarol, {\it Dark Matter in
Theories of Gauge-Mediated Supersymmetry Breaking}, CERN-TH/96-171,
hep-ph/9607225.}\
it is shown that under certain circumstances, these
particles are suitable dark matter candidates. However, the lightest of
these needs to be quite light, of order $5$ TeV (compared to a natural
scale of $30$ TeV or more). This potentially represents a fine-tuning of
one part in $30$ or worse. As these authors note,
other dark matter candidates are likely to be found elsewhere in these
models.}\ If one allows mixing, there are additional, {\it non-universal}
contributions  to scalar masses.  We will evaluate these contributions
in section 3, and find that they are negative, and proportional to the
squares of a new set of Yukawa couplings.  One might worry that, as a
result, they will spoil the good features of gauge mediation.  However,
with a minimal messenger sector, i.e. one pair of either $5+\bar5$ or
$10+\overline{10}$, only masses of one sfermion generation are shifted
and the first two generations are likely to remain degenerate.
Furthermore, we might expect that, similarly to the ordinary quark and
lepton Yukawa coupling matrix, many of these couplings are small, so only
a few states will show departures from universality.
 
In the MGM model of ref. \dnns, all supersymmetry breaking
scalar and gaugino masses depend on one parameter only. The generation
of a $\mu H_UH_D$ term in the superpotential and the generation of
a supersymmetry breaking
$BH_UH_D$ term in the scalar potential require independent mechanisms.
Furthermore, the mechanism presented in \dnns\ for generating $B$
involves fine tuning of order $(\alpha_2/\pi)^2$. It was suggested
that a discrete (possibly horizontal) symmetry
could account for the magnitude of $B$ and $\mu$, but no concrete
model was presented. In section 4 we examine the question of naturalness
in more detail. We present a specific version of the minimal model of
ref. \bkw\ where a discrete symmetry predicts $\mu$ and $B$ terms
of the correct order of magnitude.
Previous, related studies, were made in
\nref\GJP{T. Gherghetta, G. Jungman and E. Poppitz, {\it Low-Energy
Supersymmetry Breaking and Fermion Mass Hierarchies}, hep-ph/9511317.}%
\nref\DGP{G. Dvali, G.F. Giudice and A. Pomarol, {\it The Mu Problem in
Theories with Gauge Mediated Supersymmetry Breaking}, hep-ph/9603228.}%
\refs{\GJP-\DGP}.
 
One of the surprising results of this analysis is the fact that
a large $\tb$ arises naturally. In ref.
\nref\NeRa{A.E. Nelson and L. Randall, Phys. Lett. B316 (1993) 516.}%
\nref\HRS{L.J. Hall, R. Rattazzi and U. Sarid,
 Phys. Rev. D50 (1994) 7048.}%
\nref\RaSa{R. Rattazzi and U. Sarid, Phys. Rev. D53 (1996) 1553.}%
\refs{\NeRa-\RaSa}\ dimensional analysis was
used to argue that a large $\tb$ requires (in models with
two Higgs doublets) fine tuning of order $1/\tb$ in order to avoid
unacceptably light charginos. In section 5, we point out
that the existence of several energy scales in the full high energy
theory can invalidate this analysis, and in particular that it need not
hold in models of low energy supersymmetry breaking.
 
Another nice feature of the minimal messenger model of ref. \bkw,
where $A$ and $B$ vanish at tree level,
is that the supersymmetric CP violating phases, $\phi_A$ and $\phi_B$,
vanish. The supersymmetric CP problem, namely the
${\cal O}(10^{-2})$ fine tuning required in generic supersymmetric models
to satisfy constraints from electric dipole moments, is then solved.
We briefly discuss this point in section 6.
 
It is quite possible that the dynamics which breaks supersymmetry is
strongly coupled. This is an area which has only been partially explored
\ddrt. For such theories, it is more difficult to list general
constraints. We will not make a serious effort to tackle this problem
here, but we will at least enumerate some of the issues in
our concluding section, section 7.
 
%%%%%%%%%%%%%%%%%%%%%
%%%%%%%%%%%%%%%%%%%%%
\newsec{Constraints on the Messenger Sectors of Weakly Coupled Models}
 
It is possible to construct models of low energy supersymmetry breaking
using the O'Raifeartaigh and/or Fayet-Iliopoulos mechanisms, in which
all couplings are weak and which can be analyzed in perturbation theory.
One can imagine that the required couplings are small parameters
generated by some more microscopic theory. This microscopic theory might
be of the type discussed in ref. \dnns, in which dynamical supersymmetry
breaking at a not too distant scale generates such terms in an effective
action for the messengers.  Or one could imagine that it is a theory such
as string theory, and that the small mass scale is generated by
tiny non-perturbative string effects.
 
In this section, we will not worry about the detailed origin of these
terms, but instead ask about the phenomenological constraints on the
messenger sector. In a theory which is weakly coupled,\foot{This does not
necessarily mean that supersymmetry breaking arises in perturbation
theory. Models in which the hidden sector dynamics is calculable
semiclassically, for example, would fall in this class.} it is possible
to prove a number of general results.  Dimopoulos and Georgi
\ref\dg{S. Dimopoulos and H. Georgi, Nucl. Phys. B193 (1981) 150.}\
showed long ago that,
as a consequence of sum rules,
one cannot obtain a realistic spectrum at tree level
in any globally supersymmetric theory. This means that at least some
masses must be generated radiatively. In such a theory, some set of
fields, which we will call the ``messengers", must feel the breaking of
supersymmetry at tree level.  Ordinary fields will couple to these.  One
might imagine that the messengers could all be neutral under the ordinary
gauge interactions, but it is easy to rule out this possibility.
This is because only Higgs fields have the correct quantum numbers to
couple (through renormalizable interactions) to the messengers.\foot{
We are assuming here that $R$-parity is conserved, but all of the remarks
which follow are easily modified in the case of broken $R$-parity.}
But this means that Higgs masses-squared will arise at lower order (by
several loops) than gaugino masses, and so gluino masses will be far
too small.
 
So we see that the messenger sector must contain fields which are charged
under the standard model group.  These fields must come in vector-like
representations. This simply follows from the fact that these masses must
be much larger than the weak scale. If we require perturbative coupling
unification with a desert, they must come in complete $SU(5)$ multiplets.
(For a different scenario, see ref.
\ref\Alon{A.E. Farraggi, {\it Low Energy Dynamical SUSY Breaking
Motivated from Superstring Derived Unification}, hep-ph/9607296.}.)
Moreover, we can require that the couplings remain
perturbative at least up to the GUT scale. This means that one can
allow at most four $5+\bar5$'s or one $10+\overline{10}$. $SU(5)$
adjoints are not allowed.
 
Next, we must ask to what the messengers can couple. In order that they
obtain large masses, the messengers almost certainly must couple to
fields in the superpotential which obtain VEVs.\foot{Alternatively,
there might be ``bare masses" in the superpotential analogous to the
$\mu$ term. These might arise by the mechanism for the $\mu$ term
described in ref. \dnns, and further in
section 4.}  These fields must be gauge singlets. The
simplest possibility, as in the models of ref. \dnns, is that the
$F$ components of these fields also have expectation values.  These $F$
components might also arise at tree level, or through loop corrections
(e.g. mixing terms in the Kahler potential; see, for example, ref.
\ref\yanagida{T. Hotta, K.-I. Izawa and T. Yanagida,
{\it Dynamical Supersymmetry Breaking without Messenger
Gauge Interactions}, hep-ph/9606203.}).
This will lead to formulas which are simple modifications, depending on
the number of messenger fields, of eqns. \gauginomasses\ and
\scalarmasses. Alternatively, the singlets might have vanishing $F$
components. The messengers might acquire supersymmetry-breaking  masses
through loops, either involving superpotential couplings or gauge
interactions.  Aesthetic issues aside, such models will have difficulty
explaining  the $\gamma \gamma$ events,  should they turn out to be real,
since the scale of Goldstino decay constant will tend to be rather large
and the  NLSP (next to lightest supersymmetric particle)  will tend to
decay outside the detector. (This is also an issue in models in which
the $F$ component of the scalar field arises in loops.) Such models will
still lead to mass formulas somewhat different in form than those of
eqns. \gauginomasses\ and \scalarmasses. (Such models  appeared in ref.
\dns.) Of course, masses are still functions only of gauge quantum
numbers. In  this case, the number of soft breaking parameters is equal
to $8$, plus the $\mu$ and $B$ terms.
 
So it seems most likely that the messenger sector will consist of some
number, $N_5$, of $5+\bar 5$ representations ($N_5<5$), or one
$10+\overline{10}$ representation, coupling to some number, $N_S$, of
singlet fields with non-vanishing scalar and $F$ components. If there are
$N_5$ $5+\bar5$'s, and $N_S$ singlets, and we assume that there is no
mixing of the messenger fields with ordinary fields, the superpotential
in the hidden sector has the form
\eqn\generalmessenger{\sum_{i=1}^{N_S} \sum_{J,K=1}^{N_5}
(\lambda_1)_{iJK}S^i\bar q_I q_J +(\lambda_2)_{iJK} S^i\bar\ell_I\ell_K.}
For large enough $N_S$ and $N_5$, the masses of squarks, sleptons, and
gauginos, with given gauge quantum numbers, become independent
parameters. We might still expect that their masses would be arranged
hierarchically as in eqns. \gauginomasses\ and \scalarmasses,
but this is not necessarily the case.  For example, take $N_S=2$
and $N_5=1$.  Suppose that $S_1$ has a large scalar component
and a small $F$ component, while $S_2$ has a small scalar component and
a large $F$ component. Couplings to $S_1$ could give all messenger quarks
and leptons comparable supersymmetry-conserving masses, while couplings
to $S_2$ could be, say, ${\cal O}(1)$ for messenger quarks
but small for messenger leptons.  This could alter the hierarchy
(in a universal way), giving a much larger than expected ratio of squark
to slepton masses.  Similarly, one could arrange that doublets are
lighter than singlets, or that the gaugino hierarchy is altered.
 
On the other hand, the modifications of the hierarchy cannot be too
drastic, or one will face other problems. Given
the experimental constraints on squark masses, one cannot take
squarks much lighter than lepton doublets, without having to fine tune
Higgs parameters.  Similarly, if squarks are extremely heavy,
one will have an extremely large, negative contribution
to Higgs masses, and further problems with fine tuning.
Finally, if one wants to explain the $\gamma\gamma\not E_T$
events in this framework, the fundamental scale of supersymmetry breaking
cannot be much larger than $10^3-10^4$ TeV.
Still, it is worth keeping in mind
that the hierarchy of squark and gaugino masses suggested by the MGM need
not hold, even in weakly coupled theories, provided that they are
sufficiently complicated. It is a simple matter to perform the analogous
analysis when the messenger sector contains a single $10+\overline{10}$.
 
So far, we have explained how modifications of the hierarchy
might arise, but not violations of universality.
The fields $\bar q$ have the same quantum numbers as the ordinary
$\bar d$ fields. We define $\bar d$ as the three fields that do
not have a couplings of the form \simplemodel\ or \generalmessenger.
In the models of ref. \dnns\ (and most other recent works), it was
implicitly assumed that only the $\bar d$ fields have Yukawa couplings,
$H_D Q\bar d$. Indeed, terms of the form $H_D Q\bar q$ can be forbidden
by discrete symmetries. (Analogous comments hold for the lepton fields or
for $10+\overline{10}$ messengers.)
On the other hand, such Yukawa couplings may be present, and can lead to
more profound modifications of the minimal gauge mediated theory than we
have contemplated up to now. Moreover, in the absence of these couplings,
the messenger sector contains stable or nearly stable particles which may
be problematic in cosmology.  In the next section, we explore the
consequences of introducing such couplings.
 
%%%%%%%%%%%%%%%%%%%%%%%%%%%%%%%%%%%
%%%%%%%%%%%%%%%%%%%%%%%%%%%%%%%%%%%
\newsec{Messenger--Matter Mixing}
 
In order to understand possible modifications of the spectrum
in the presence of mixing between messenger fields and
ordinary matter fields, consider first the model of eqn. \simplemodel.
The VEV $\vev{S}$ gives a supersymmetric contribution to the mass of the
messenger quarks and leptons, while $\vev{F_S}$ leads to a
supersymmetry-violating splitting
in these multiplets.  At one loop, gauginos gain mass through their
couplings to these fields; at two loops, ordinary squarks and sleptons
gain mass.  In eqns. \gauginomasses\ and \scalarmasses, the
parameter $\Lambda$ is given by $\Lambda=F_S/S$ (here and
below, expectation values are understood).
 
The simple modification that we consider takes place in the Yukawa
sector. The messenger $\ell$ field has the same gauge quantum numbers
as the ordinary lepton doublets; $\bar q$ has the same
quantum numbers as the $\bar d$ quarks.  Thus, in the absence
of a symmetry, one expects these fields to mix. In particular,
in the Yukawa couplings,
\eqn\yukawa{H_D L_i Y^{\ell}_{ij}\bar e_j+H_DQ_i Y^d_{ij}\bar d_j,}
each of $L_i$ and $\bar d_i$ refers to the four objects
with the same quantum numbers. Then $Y^\ell$ is a $4\times3$ matrix
while $Y^d$ is a $3\times4$ matrix. By convention, we call
$L_4$ and $\bar d_4$ the linear combination of fields which
couple to $S$ in eqn. \simplemodel. We refer to $Y^\ell_{4i}$
and $Y^d_{i4}$ as {\it exotic} Yukawa couplings.
 
The exotic Yukawa couplings contribute, through one-loop diagrams, to the
masses of the ordinary squarks and sleptons. These diagrams are indicated
in fig. 1.
 
\epsfbox{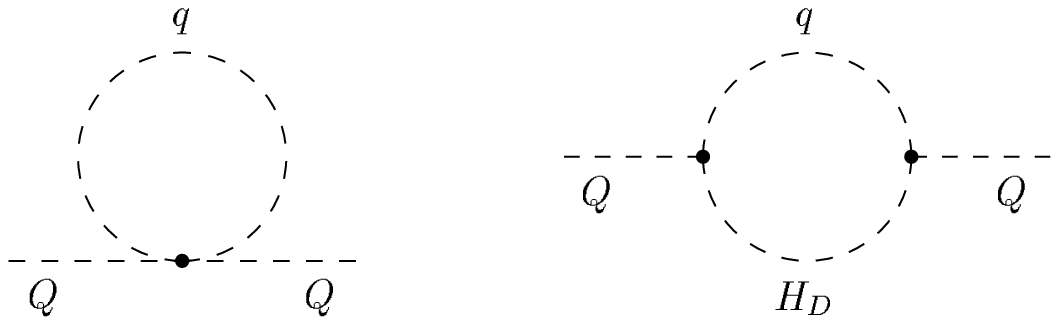}
\vbox{%
{\narrower\noindent%
\multiply\baselineskip by 3%
\divide\baselineskip by 4%
{\rm Figure 1. }{Scalar-loop contributions to squark mass shifts. $Q$ are
ordinary left-handed squark doublets, $H_D$ is the down Higgs doublet,
and $q$ are the messenger squarks.
\medskip}}}
%\fig\figA{Scalar-loop contributions to squark mass shifts. $Q$ are
%ordinary left-handed squark doublets, $H_D$ is the down Higgs doublet,
%and $q$ are the messenger squarks.}.
 
It is a simple matter to compute these in a power series in
$F_S/S^2$. The zeroth order term, of course, vanishes by
supersymmetry. The first order term vanishes as a result of an accidental
cancellation. In order to understand the result, suppose first that only
one of the sleptons, say $\bar e_3$, has a substantial Yukawa coupling to
$L_4$ and call this coupling $y_\ell$. (There is
actually no loss of generality
here. In general, the affected slepton is the combination
$\sum_i Y^\ell_{4i}\bar e_i$, with $y_\ell^2=\sum_i(Y^\ell_{4i})^2$.)
The mass shift is
\eqn\massshift{\delta m_{\bar e_3}^2=-{M^2 \over 6}{y_\ell^2\over16\pi^2}
{\vert F_S \vert^4 \over M^8},}
where $M=\lambda_2 S$ is the mean mass of the $\ell_4$ multiplet.
Using eqn. \scalarmasses\ for the (universal) two loop contribution to
$m^2_{\bar e}$, we find
\eqn\onetwoE{
{\delta m_{\bar e_3}^2\over m_{\bar e}^2}=-{1\over12}
{y_\ell^2\over\alpha_Y^2}{|F_S|^2\over M^4}
\approx-10^3\ y_\ell^2\ {\vert F_S \vert^2 \over M^4}.}
There is also a related shift of the down Higgs mass:
\eqn\onetwoH{{\delta m_{H_D}^2\over m_{H_D}^2}
=-{1\over9}{y_\ell^2\over\alpha_2^2}{|F_S|^2\over M^4}
\approx-10^2\ y_\ell^2\ {\vert F_S \vert^2 \over M^4}.}
 
A few comments are in order, regarding the results \onetwoE\ and
\onetwoH:
\item{(i)} Since the result of eqn. \massshift\ is proportional to
$|F_S|^4$, in contrast to the two loop contribution of eqn.
\scalarmasses\ which is proportional to $|F_S|^2$, there is a
natural way of understanding how, even for Yukawa couplings of order one,
one loop corrections could be comparable to two loop gauge corrections,
rather than much larger. Explicitly, $\delta m_{\bar e_3}^2/m_{\bar e}^2
\lsim1$ if $|F_S|/M^2\lsim0.03$.
\item{(ii)}  Related to (i), it is important that contributions to
masses of squarks and sleptons not be too large, or charged
or colored fields will obtain expectation values. For our
example above, the negative correction to the Higgs mass is of the same
order.  But given that the correction to the singlet cannot be too large,
the fractional correction to the doublet mass will be rather small.
\item{(iii)} With a single messenger $5+\bar5$ pair, this mass shift
affects only one right-handed slepton generation. The other two
remain degenerate.
\item{(iv)} If, similarly to ordinary Yukawa couplings,
$Y^{\ell}_{4\tau}\gg Y^\ell_{4i}$ for $i=e,\mu$, then the shift
is in the mass of the right-handed stau. The degeneracy of the
selectron and smuon guarantees that all constraints from
flavor changing neutral processes are satisfied.
 
In this simple model, there is also a shift in the mass
of one of the left-handed squark doublets,
\eqn\onetwoQ{{\delta m_{Q_3}^2\over m_Q^2}
=-{1\over16}{y_d^2\over\alpha_3^2}{|F_S|^2\over M^4}
\approx-6\ y_d^2\ {\vert F_S \vert^2 \over M^4},}
(where $y_d^2=\sum_i(Y^d_{i4})^2$) and a related shift
in the mass of the down Higgs, so that \onetwoH\ is modified to
\eqn\onetwoHa{{\delta m_{H_D}^2\over m_{H_D}^2}
=-{1\over9}{3y_d^2+y_\ell^2\over\alpha_2^2}{|F_S|^2\over M^4}
\approx-10^2\ (3y_d^2+y_\ell^2)\ {\vert F_S \vert^2 \over M^4}.}
Again, if the exotic Yukawa coupling is largest for $Q_3$, then
$Q_1$ and $Q_2$ remain degenerate and constraints from flavor changing
neutral processes (e.g. $K-\bar K$ and $D-\bar D$ mixing)
are easily satisfied. If we adopt $|F_S|/M^2\lsim0.03$, then both
\onetwoQ\ and \onetwoHa\ are small.
 
The most plausible effect of the mixing is then a (negative)
shift in the mass of $\tilde\tau_R$. There is a small shift in the
squared-mass of $H_D$, while for all other scalars, the one-loop
mixing contribution is either absent or very small.
It is possible, however, that $y_\ell\ll y_d\lsim1$.
In this case, a substantial shift in $m^2_{H_D}$ is possible
with a corresponding (but much less substantial) shift in
$m^2_{Q_3}$. Finally, if the generation hierarchy of the
exotic Yukawa couplings is very different from the ordinary
Yukawa couplings, the result could be that, say, the selectron or the
smuon is the lightest among the right-handed sleptons. But then the
constraints from $\mu\ra e\gamma$ are significant and require that the
splitting is small. Similarly, constraints from $K-\bar K$ mixing require
that the splitting in the squark sector is small if the largest exotic
Yukawa coupling is $Y^d_{14}$ or $Y^d_{24}$.
 
Next, consider models with $N_5>1$. Here, for generic mixing between
messenger and matter fields, all three generations of left-handed
squarks and of right-handed sleptons are split. Flavor changing neutral
current constraints are significant. But if we take, as above,
$|F_S|/M^2\lsim0.03$ and, in addition, assume that the exotic Yukawa
couplings are not larger than the corresponding ordinary Yukawa
couplings, e.g. $Y^\ell_{4\mu}\lsim m_\mu\tan\beta/m_t$, then
all the constraints are satisfied.
Such a hierarchy in the exotic
Yukawa couplings is very likely if the smallness and hierarchy of
the ordinary Yukawa couplings is explained by horizontal symmetries
(see, for example,
\ref\LNS{M. Leurer, Y. Nir and N. Seiberg,
 Nucl. Phys. B420 (1994) 468.}).
 
Finally, we may consider mixing with messenger $10+\overline{10}$.
Then masses of all ordinary scalar fields, except for the right-handed
sleptons, are shifted. Again, for a single pair of $10+\overline{10}$,
only one generation in each sector is affected. Very plausibly, these are
the third generation sfermions, so that constraints from flavor changing
neutral current processes are rather weak. A small parameter $|F_S|/M^2$
guarantees that these one loop corrections are smaller than or comparable
to the two loop gauge contributions. Substantial corrections could occur
for the slepton and Higgs fields, but the mass shifts for squarks
are small.
 
We learn then that there are a few possibilities concerning
the effects of messenger--matter mixing:
\item{a.} There is no mixing or the mixing is negligibly small.
Eqns. \gauginomasses\ and \scalarmasses\ remain valid.
This is the situation if there is a symmetry that forbids mixing
or if the ratio $|F_S|/M^2$ is small.
\item{b.} There is a large negative mass shift of order one
for $\tilde\tau_R$ and a small negative mass shift of order 0.1
for $H_D$. For all other soft supersymmetry breaking parameters,
\gauginomasses\ and \scalarmasses\ remain an excellent approximation.
This is the situation if $y_\ell^2|F_S/M^2|\sim0.03$
\item{c.} There is a large negative mass shift of order one
for $H_D$ and a small negative mass shift of order $0.02$ for
$Q_3$. For all other soft supersymmetry breaking parameters,
\gauginomasses\ and \scalarmasses\ remain an excellent approximation.
This is the situation if $y_d^2|F_S/M^2|\sim0.06$ and $y_\ell\ll y_d$.
\item{d.} The lightest squark or slepton could belong to the
first or second generation or all three generations could be
split in masses. This is the situation if the hierarchy in the
exotic Yukawa couplings is different from that of the ordinary ones
or if there are several $5+\bar5$ representations. But then
phenomenological constraints require that the mass shifts are small.
 
We emphasize that the effects cannot be large in the squark sector.
But there could be large effects in the slepton and/or Higgs sectors.
Such corrections might be helpful in understanding at least one issue.
In low energy breaking, there are potential fine-tuning problems in
obtaining a suitable breaking of $SU(2) \times U(1)$. The problem is that
the masses of the lightest right-handed leptons are constrained,
from experiment, to be greater than about $45$ GeV.
On the other hand, if gauge mediation is the
principle source of all masses, the contribution to the masses of the
Higgs doublets tends to be larger.  So if the lightest slepton
has a mass of order $80$ GeV or more (as suggested by the
CDF event) then the typical contributions to Higgs masses
would seem to be on the large side.  Additional negative
contributions would tend to ameliorate this problem.

Finally, we should mention on other possible source of violation of
universality. Throughout this discussion, we have assumed an underlying
$R$-parity, and that $q$ and $\bar q$ have the same $R$ parity
as ordinary quarks. It is possible that $R$ parity is broken, or that
$q$ and $\bar q$ have the {\it opposite} R parity. In this case,
operators like $LQ\bar q$ or $\bar u\bar d\bar q$ may be allowed.
The latter can lead to more appreciable shifts in squark masses, and
thus more significant violations of universality in the squark sector
then we have contemplated up to now.
 
%%%%%%%%%%%%%%%%%%%%%%%%%%
%%%%%%%%%%%%%%%%%%%%%%%%%%
\newsec{The $\mu$ Problem}
 
In the MGM model of ref. \dnns, the following mechanism to generate
a $\mu$ term was employed. An additional singlet field $T$ was
introduced, which couples to the Higgs fields through a nonrenormalizable
term in the superpotential,
\eqn\THH{{T^n\over M^{n-1}}H_UH_D.}
To generate a $B$ term, it was suggested that a term in the
superpotential of the form
\eqn\SHH{\lambda_h S H_UH_D}
is allowed. With a small $\lambda_h\sim(\alpha_2/\pi)^2$, it gives an
acceptable $B\sim(\alpha_2/\pi)^2 F_S$ and a negligible contribution
to $\mu$. It is difficult, if not impossible, to find a symmetry
that forbids all $H_UH_D$ couplings except for \THH\ and -- with
an appropriately small $\lambda_h$ -- \SHH. (For previous, unsuccessful
attempts, see \GJP.) If, however, \SHH\ is forbidden or highly
suppressed, so that $B=0$ at tree level, then loop contributions still
generate $B\sim(\alpha_2/\pi)^2\Lambda\mu$ \bkw, which
is small but not negligibly small. We now present a simple model where,
indeed, as a result of a discrete symmetry, \THH\ gives the largest
contribution to $\mu$ while $\lambda_h$ of \SHH\ is negligibly small.
As we will explain in section 6, such a model offers hope of solving
the supersymmetric CP problem.
 
Let us introduce a (horizontal) symmetry $H=Z_m$ and and set the
$H$-charges of the relevant fields to
\eqn\Hhiggs{H(S)=0,\ \ \ \ H(H_UH_D)=n,\ \ \ \ H(T)=-1.}
The various VEVs are hierarchical, $\vev{T}\gg\sqrt{\vev{F_S}}\gg
\vev{H_U},\vev{H_D}$ and spontaneously break, respectively, the symmetry
$H$, supersymmetry (and an $R$ symmetry) and the electroweak symmetry.
The relevant terms in the superpotential are
\eqn\SuPo{\eqalign{W=&\ W_0(S)+W_1(S,T)+W_2(S,T,H_U,H_D),\cr
W_1\sim&\ {T^m\over M_p^{m-3}}\left(1+{S\over M_p}+\cdots\right),\cr
W_2\sim&\ {T^nH_UH_D\over M_p^{n-1}}\left(1+{S\over M_p}+\cdots\right).
\cr}}
Here, $M_p$ is the Planck scale which suppresses all nonrenormalizable
terms. The dots stand for terms that are higher order in $S/M_p$.
 
Similarly to the model of Abelian horizontal symmetries presented in
\LNS, the minimum equations give an $H$-breaking scale that is
intermediate between the supersymmetry breaking scale and the Planck
scale and
depends only on $m$. Explicitly, ${\partial V\over\partial T}=0$ gives
\eqn\mineq{{F_S\over M_p^{2}}\sim\left({T\over M_p}\right)^{m-2}.}
 
Also similarly to the models of \LNS, the supersymmetric $\mu$ problem
is solved because a term $\mu H_U H_D$ violates $H$.
The leading contribution to $\mu$ is of order
\eqn\muH{{\mu\over M_p}\sim\left({T\over M_p}\right)^n\sim
\left({F_S\over M_p^2}\right)^{n\over m-2}.}
For definiteness, we take $F_S/M_p^2\sim10^{-28}$ and require that \muH\
predicts $\mu/M_p\sim10^{-16}$. This is the case for
$n\approx{4\over7}(m-2)$. The simplest option is then $n=4$ and $m=9$
(corresponding to $T/M_p\sim10^{-4}$). If one insists on larger $T/M_p$,
so that it may be relevant to the fermion mass hierarchy, say $10^{-3}$
($10^{-2}$), it can be achieved with $n=5,m=11$ ($n=8,m=16$).
 
A $B$ term is also generated by $W$ of eq. \SuPo. The leading
contribution is of order
\eqn\BH{B\sim {F_S\mu\over M_P}.}
This contribution to $B$ is $\ll\mu^2$ and, therefore, negligible.
A much larger contribution is generated at the two loop level
(note that $M_2$ is generated by one loop diagrams):
\eqn\Btwo{B\sim{\alpha_2\over\pi}\mu M_2.}
This is smaller that the square of the electroweak symmetry breaking
scale by a factor of order $\alpha_2$. Consequently, $\tan\beta$
is large, of order $\alpha_2^{-1}$.
 
%%%%%%%%%%%%%%%%%%%%%%%%%%%%%%%%
%%%%%%%%%%%%%%%%%%%%%%%%%%%%%%%%
\newsec{Naturally Large $\tb$}
 
It has been argued \refs{\NeRa-\RaSa}\ that, if there are only two
Higgs doublets in the low energy supersymmetric model, large
$\tb$ requires a fine tuning in the parameters of the Lagrangian of order
$(1/\tb)$. The naturalness criterion used, for example, in ref.
\NeRa\ states that {\it ``unless constrained by additional approximate
symmetries, all mass parameters are about the same size, and all
dimensionless numbers are of order one."} However, in all existing models
of dynamical suppersymmetry breaking (DSB), there is more than one
relevant energy scale. The assumption that all dimensionful parameters
are characterized by a single scale may fail. Then large $\tb$ may arise
naturally, as is the case in the model of the previous section.
 
Let us first repeat the argument that large $\tb$ requires fine tuning.
The basic assumption here is that, in the low energy effective
supersymmetric Standard Model, there is a single scale, that is the
electroweak (or, equivalently, the supersymmetry) breaking scale.
A dimensionful parameter can be much smaller only as a result of an
approximate symmetry. The Higgs potential for the two Higgs doublets is
\eqn\Higgspot{m_U^2H_U^2+m_D^2H_D^2+B(H_UH_D+{\rm h.c.})
+{g^2+g^{\prime2}\over8}(|H_U|^2-|H_D|^2)^2.}
In the large $\tb$ region,
\eqn\largetb{{1\over\tb}\approx-{B\over m_U^2+m_D^2}.}
Large $\tb$ requires $B\ll m_U^2+m_D^2$. There are two symmetries
that could suppress $B$ below its natural value of order $m_Z^2$.
If $B$ is made small ($B\sim m_Z^2/\tb$) by an approximate R symmetry,
the wino mass $M_2$ should also be small $(M_2\sim m_Z/\tb$).
If $B$ is made small by an approximate PQ symmetry, then the $\mu$ term
should also be small ($\mu\sim m_Z/\tb$).
This has interesting consequences for the chargino mass matrix,
\eqn\chargino{\pmatrix{\mu&{g\over\sqrt2}\vev{H_U}\cr
{g\over\sqrt2}\vev{H_D}&M_2\cr}.}
As $\vev{H_D}$ is small by assumption and as (to make $B$ naturally
small) at least one of $\mu$ and $M_2$ has to be small, the mass matrix
\chargino\ leads to a light chargino (with mass of order $m_Z/\tb$).
This is phenomenologically unacceptable (the
bounds on chargino masses are roughly $\gsim m_Z/2$).
This means that the natural scale for either $\mu$ or $M_2$ is
of order $m_Z\tan\beta$, and the criterion for naturalness is violated.
 
The assumption that a natural effective low energy supersymmetric
Standard Model has a single energy scale is a strong
one. In all existing models of DSB, there are at least
three energy scales: the Planck scale $M_p$, the supersymmetry
breaking scale $M_S$, and the electroweak breaking scale $m_Z$.
Whether indeed $m_Z$ is the only relevant scale for the low energy
theory and, in particular, for $\mu$ and $B$, is a model dependent
question. In hidden sector models of supersymmetry breaking, one assumes
that $m_Z\sim M_S^2/M_p$ is, indeed, the only relevant scale
in the low energy model. But this is a rather arbitrary (though
convenient) ansatz and, in the absence of a detailed high energy
theory for the messenger sector, does not stand on particularly
firm grounds. The situation is even more complicated in models
of gauge mediated supersymmetry breaking.
Here, in addition to the Planck scale, there exist the dynamical
supersymmetry breaking scale $\sqrt{F_S}$, the scale
$\Lambda=F_S/S$ and the electroweak scale $m_Z\sim\alpha_2\Lambda$.
Which of these scales is relevant to $B$ depends on the mechanism
that generates $B$. It could very well be that the natural scale
for $B$ is $B \ll m_Z$.
 
To understand the situation in more detail, let us assume
that there is neither a PQ symmetry nor an R symmetry to suppress $B$.
Then the natural value for $\mu$ is $M_p$, and the model does
not provide any understanding of the $\mu$ problem. But even if we
assume that $\mu\ll M_p$ for some reason,
the term \SHH\ is allowed. This leads to $\mu\sim S$ and $B\sim F_S$.
Both values are unacceptably large, but our main point here is that
the natural scale for $B$ could easily be the {\it highest}
supersymmetry breaking scale in the {\it full} theory, $M_S$.
 
In the model presented in the previous section, the $H$ symmetry
leads to an accidental PQ symmetry. The small breaking parameter
of the PQ symmetry is of order $m_Z/M_p$, thus solving the
$\mu$ problem. At the same time, it leads to a tree level value
for $B$ that is of order $F_S\mu/M_p$. This is actually similar
to the scale in supergravity models, except that in those models
$F_S\mu/M_p\sim m_Z^2$ while in models of GMSB
$F_S\mu/M_p\ll m_Z^2$. Consequently, this contribution to
$B$ is negligibly small. The main point here is that the natural
scale for $B$ could  be $M_S^2/M_p$; this scale
coincides with the electroweak scale $m_Z$ only in supergravity models.
 
Finally, a larger contribution to $B$ arises in our model from
two loop diagrams, of order $\alpha_2\mu M_2\sim\alpha_2^2\mu\Lambda$.
This is smaller than the electroweak scale, $m_Z^2\sim\alpha_2^2
\Lambda^2$ by a factor of order $\mu/\Lambda\sim\alpha_2$. We learn that
different combinations of scales could be relevant to
$B$ and to $m_Z$. If the combinations are such that $B\ll m_Z^2$,
then a large $\tb$ arises and no fine tuning is required.
The model of the previous section provides a specific example
of this situation.
 
%%%%%%%%%%%%%%%%%%%%%%%%%%%%%%%%
%%%%%%%%%%%%%%%%%%%%%%%%%%%%%%%%
\newsec{The Supersymmetric CP Problem}
 
Supersymmetric theories introduce new sources of CP violation.
With the minimal supersymmetric extension of the Standard Model
and assuming universality of gaugino and of sfermion masses,
there are four additional phases beyond
the Kobayashi-Maskawa phase and $\theta_{\rm QCD}$ of the Standard
Model. One phase appears in the $\mu$ parameter, and the other
three in the soft supersymmetry breaking parameters $M_\lambda$,
$A$ and $B$,
\eqn\Lag{{\cal L}=-{1\over2}M_\lambda\lambda\lambda-A(h_uQH_U\bar u
-h_d QH_D\bar d-h_\ell LH_d\bar e)-BH_UH_D+{\rm h.c.},}
where $\lambda$ are the gauginos and $h_i$ the Yukawa couplings.
Only two combinations of the four phases are physical
\nref\DGH{M. Dugan, B. Grinstein and L. Hall,
 Nucl. Phys. B255 (1985) 413.}%
\nref\DiTh{S. Dimopoulos and S. Thomas,
 Nucl. Phys. B465 (1996) 23, hep-ph/9510220.}%
\refs{\DGH-\DiTh}. These can be taken to be
\eqn\physical{\eqalign{
\phi_A=&\ \arg(A^*M_\lambda),\cr
\phi_B=&\ \arg(B\mu^* M_\lambda^*).\cr}}
Unless these phases are $\lsim{\cal O}(10^{-2})$, or supersymmetric
masses are $\gsim{\cal O}(1\ TeV)$, the supersymmetric contribution
to the electric dipole moment of the neutron is well above the
experimental bound. This is the supersymmetric CP problem.
 
In models of GMSB, gaugino masses are not universal (see eq.
\gauginomasses). However, with a minimal messenger sector
($N_S=N_5=1$), gaugino masses carry a universal phase. Thus, there
still exist only the two new phases defined in eqn. \physical.
 
In the MGM model of ref. \dnns, $A(\Lambda)=0$. In its minimal version
investigated in ref. \bkw, also $B(\Lambda)=0$. Radiative corrections
give \bkw:
\eqn\radAB{\eqalign{
A_t\simeq&\ A_q(\Lambda)+M_2(\Lambda)\left[-1.85+0.34|h_t|^2\right],\cr
{B\over\mu}\simeq&\ {B\over\mu}(\Lambda)-{1\over2}A_t(\Lambda)
+M_2(\Lambda)\left[-0.12+0.17|h_t|^2\right].\cr}}
Using \physical, we learn from \radAB\ that, for $A(\Lambda)=
B(\Lambda)=0$, one has
\eqn\MGMAB{\phi_A=\phi_B=0.}
Thus, the supersymmetric CP problem is solved in this model.
 
The vanishing of the supersymmetric phases goes beyond the
approximation \radAB. It is actually common to all models
with universal sfermion masses and a universal phase in
the gaugino masses and where, at tree level, $A=B=0$.\foot{
We thank Riccardo Rattazzi for explaining this point to us.}
In the absence of non-gauge interactions, there is an additional
$R$ symmetry in the supersymmetric Standard Model. In a spurion
analysis, it is possible to assign the same $R$ charge to $M_{\lambda}$,
$A$ and $B$
\nref\HRS{L.J. Hall, R. Rattazzi and U. Sarid,
 Phys. Rev. D50 (1994) 7048, hep-ph/9306309.}%
\refs{\HRS,\DiTh}. If the only source of $R$ symmetry breaking
are gaugino masses, both $\phi_A$ and $\phi_B$ are zero, just because
$A$, $B$ and the gaugino mass have the same $R$ charge, and the
RG evolution formally respects the $R$ symmetry.
 
At the two-loop level, Yukawa interactions affect the running of
$A$. Proportionality of the $A$-terms and the Yukawa terms is
violated and complex phases (related to the Kobayashi-Maskawa
phase) appear in off-diagonal $A$-terms (see refs.
\nref\BBMR{S. Bertolini, F. Borzumati, A. Masiero and G. Ridolfi,
 Nucl. Phys. B353 (1991) 591.}%
\nref\BeVi{S. Bertolini and F. Vissani, Phys. Lett. B324 (1994) 164.}%
\refs{\BBMR-\BeVi}\ for the relevant RGE). The contribution of these
phases to the electric dipole moment of the neutron is, however,
highly suppressed.
 
We conclude then that in the minimal version of MGM models
(namely when $A=B=0$ at a high scale) the supersymmetric
CP problem is solved.
 
%%%%%%%%%%%%%%%%%%
%%%%%%%%%%%%%%%%%%
\newsec{Conclusions}
 
If more events with two photons plus missing energy are discovered,
this can be viewed as strong evidence for low energy supersymmetry
breaking. The MGM has a strong appeal, given its simplicity, but
one can easily imagine that the messenger sector may be more complicated.
It is possible that the data will support the MGM, but even given the
limited information we have now, there are hints that some extension of
the model may be required \dtw. We have seen that in weakly coupled
theories the spectrum can be modified in two significant ways.
First, the hierarchy may be altered.  As a result, one can imagine that,
say, slepton doublets are not much more massive than singlets (as
suggested in ref. \dtw). Second, there can be departures from
universality.  In other words, some $SU(2)$ singlet sleptons might be
lighter than others. We have seen that there are significant constraints
on such universality violations coming, for example, from requiring
reasonable breaking of $SU(2) \times U(1)$. We have also seen that if
horizontal symmetries are responsible for the hierarchies of ordinary
quark and lepton masses, at most only a few states will exhibit
appreciable universality violation (e.g. the stau may be significantly
lighter than the other sleptons).
 
We have so far avoided the more difficult question of what
may happen in strongly coupled theories.  These issues were
touched upon in \ddrt.  In the event that the underlying
supersymmetry breaking theory is strongly coupled, it seems
likely that some of our constraints will be relaxed.  For example,
it is not clear that asymptotic freedom is a correct criterion,
since we know from the work of Seiberg
\ref\Seib{N. Seiberg, Nucl. Phys. B435 (1995) 129, hep-ph/9411149.}\
that the infrared degrees of freedom of a theory may be quite different
than the microscopic degrees of freedom.  Another difficulty lies in
mass formulas such as eqns. \scalarmasses\ and \gauginomasses.
It is not clear whether in strongly coupled
theories, the factors of $(4 \pi)^{-2}$ which appear in weak coupling
will also appear.  For example, there may be single particle
states which can appear in two point function relevant
to the gaugino mass computation, and one might
suspect that the result, lacking the usual phase space factors,
will be larger.  Thus one can imagine that the susy breaking
scale might be closer than suggested by weak coupling
models.  This possibility should be taken seriously, since one might hope
in such a framework to avoid  the division into different sectors which
we have seen is inevitable in weakly coupled models.
 
We have also discussed the $\mu$ problem and the question of large
$\tan\beta$. We have noted that the usual arguments that large
$\tan\beta$ requires fine tuning make assumptions about the scales $\mu$
and $B$ which need not hold -- indeed one might argue are not likely
to hold -- in theories of low energy dynamical breaking. In particular,
it is quite natural for $B$ to be very small at the high scale \bkw.
In this situation, the supersymmetric CP problem is automatically solved.
 
The MGM models are attractive in that they are highly predictive,
guarantee universality, can suppress the supersymmetric CP violating
phases, and predict events with final photons and missing energy
similar to the one observed by CDF. In this work we have learned that
reasonable extensions of the minimal models retain many of these
nice features while offering a richer phenomenology:
\item{a.} The number of parameters describing sfermion and gaugino
masses can increase to eight with extended messenger sectors, or to
about eleven with messenger - matter mixing. The hierarchy of masses
between, say, gauginos and sfermions or squarks and sleptons may be
different from the minimal models.
\item{b.} Universality is violated with messenger - matter mixing
but, most likely, it is only the third generation that is significantly
affected. Interesting flavor changing neutral current
processes may be observed, for example, in tau decays.
\item{c.} Final photons and missing energy remain the typical
signature of low energy supersymmetry breaking, but the
detailed nature of the final states could be rather different
than in the MGM models.

%%%%%%%%%%%%%%%%%%%%%%%%%%%%%
 
\vskip 1cm
{\bf Acknowledgments:}
We thank Savas Dimopoulos, Scott Thomas, Francesca Borzumati and
Riccardo Rattazzi for important comments and criticisms as these
ideas were developed.
Y.N. is supported in part by the United States -- Israel Binational
Science Foundation (BSF), by the Israel Commission for Basic Research,
and by the Minerva Foundation (Munich).  The work of M.D. is supported
in part by the U.S. Department of Energy.
 
\listrefs
%\listfigs
 
%\epsfbox{dig2.eps}
 
\bye